# "ASYMPTOTIC PARABOLA" FITS FOR SMOOTHING GENERALLY ASYMMETRIC LIGHT CURVES

K.D. Andrych [1], I.L. Andronov [2], L.L. Chinarova [1], V.I. Marsakova [1]

[1] Department of Astronomy and Astronomical Observatory, Odessa National University, Marazlievskaya 1V, 65014, Odessa, Ukraine, *katyaandrych@gmail.com, lidia_chinarova@mail.ru, v.marsakova@onu.edu.ua*

[2] Department "High and Applied Mathematics, Odessa National Maritime University, Mechnikova Str., 34, 65029, Odessa, Ukraine, *tt_ari @ ukr.net*

**ABSTRACT.** A computer program is introduced, which allows to determine statistically optimal approximation using the "Asymptotic Parabola" fit, or, in other words, the spline consisting of polynomials of order 1,2,1, or two lines ("asymptotes") connected with a parabola. The function itself and its derivative is continuous. There are 5 parameters: two points, where a line switches to a parabola and vice versa, the slopes of the line and the curvature of the parabola. Extreme cases are either the parabola without lines (i.e. the parabola of width of the whole interval), or lines without a parabola (zero width of the parabola), or "line+parabola" without a second line. Such an approximation is especially effective for pulsating variables, for which the slopes of the ascending and descending branches are generally different, so the maxima and minima have asymmetric shapes. The method was initially introduced by Marsakova and Andronov (1996OAP.....9..127M) and realized as a computer program written in QBasic under DOS. It was used for dozens of variable stars, particularly, for the catalogs of the individual characteristics of pulsations of the Mira (1998OAP....11...79M) and semi-regular (200OAP....13..116C) pulsating variables. For the eclipsing variables with nearly symmetric shapes of the minima, we use a "symmetric" version of the "Asymptotic parabola". Here we introduce a Windows-based program, which does not have DOS limitation for the memory (number of observations) and screen resolution. The program has an user-friendly interface and is illustrated by an application to the test signal and to the pulsating variable AC Her.

## 1. Introduction

Variable stars are important sources of information on structure and evolution of stars, They are observed by many professional and amateur astronomers, which have generally different photometrical systems. In this case, it is not easy to reduce all the data to some effective system, thus such observations are used for high-amplitude variables, e.g. Mira – type stars, cataclysmic binaries or semi – regular variables. A huge part of observations of such objects was obtained by amateurs and stored in national (but really international) databases like AAVSO (http://aavso.org), AFOEF (http://cdsarc.u-strasbg.fr/afoev/), VSOLJ (http://vsolj.cetus-net.org/). Photographical surveys ("Sky Patrol") produced "one or few per night" observations, which are not suitable for short – period (or aperiodic stars with a short time – scale of variability). The most abundant collections of photo negatives are stored in the AAVSO (Harvard Observatory), Sonneberg observatory, and Astronomical Observatory of the Odessa National University. Nowadays, there are numerous "photometric surveys" based on the CCD observations from ground-based and space telescopes. They are very important in studying variable stars (including the newly discovered ones) "in the past".

However, for eclipsing and pulsating variables, there is another very popular mode of observations: time series, which are obtained during a time interval, which is shorter than a photometric period – close to the minimum of eclipsing binaries, or to the maximum of pulsating variables. Then from the light curve is determined a single parameter – the moment of the extremum and its accuracy (needed, but often ignored). These individual extrema are published in the journals "Variable Stars. Supplement" ("Peremennye Zvezy. Prilozhenie"), "Information Bulletion on Variable Stars", "Open European Journal on Variable Stars" et al. and are used for studies of the period changes using the method of the "O-C" diagrams (see Tsessevich 1971 for more details).

The methods for determination of the extrema were based on the Pogson's method of chords, Hertzsprung's method of fitting the data to some "standard curve", method of "tracing paper" (see e.g. Tsessevich (1971) for a review). Kwee and van der Woerden (1956) proposed a very popular method, which was implemented in some computer programs. However, it generally produces unreal small error estimates and thus is to be replaced by fitting the data with some approximating function (e.g. Andronov 1994, 2005, Andronov and Marsakova 2006, Mikulášek 2007).

These methods are based on the assumptions of "symmetry" (suitable for eclipses of binary stars) and "asymmetry" (suitable for maxima of pulsating and cataclysmic variables). For latter, Marsakova and Andronov (1996) proposed the method of "asymptotic parabolae"), which was successfully used for determination of characteristics of thousands of maxima and minima of individual cycles of variability of single and binary stars of different types (Marsakova and Andronov 1998)..

As in the electronic tables (Microsoft Excel, Open/Libre Office, GNUmeric), the smoothing functions are algebraic polynomials. The polynomials were applied for the determination of more than 6500 extrema (and the error estimates) of semi-regular variables by Chinarova and Andronov (2000).

The method was also realized later in the programs VSCalc (Breus 2007) and PERANSO (http://peranso.com)

**2. Basic Equations**

In this work, we present application of algebraic polynomials and asymptotic parabolae for determination of extrema. Following Andronov (1994, 2005), we make a mathematical model (smoothing function) for the signal:

$$x_C(t) = \sum_{\alpha=1}^{m} C_\alpha f_\alpha(t) \qquad (1)$$

Here $C_\alpha$ are called "the coefficients" and $f_\alpha(t)$ - "the basic functions". The coefficients $C_\alpha$ are usually determined by minimizing the test function

$$\Phi_m = \sum_{jk=1}^{n} h_{jk}(x_j - x_{Cj})(x_k - x_{Ck}), \qquad (2)$$

which uses the observed $x_j$ and computed $x_{Cj} = x_C(t_j)$ values. Obviously, results are dependent on the matrix $h_{jk}$, which is similar to the "metric tensor" in the differential geometry. In the electronic tables, the matrix $h_{jk}$ is oversimplified to the unitary matrix (Kronecker symbol) $\delta_{jk}$ (equal to 1, if $j = k$, and else 0). More general case is $h_{jk} = w_k \delta_{jk}$, $w_k = \sigma_0^2 / \sigma_k^2$, $\sigma_0$ – the "unit weight error" and $\sigma_k$ - the accuracy of the observation $x_k$.

For the derivative of degree $s$, the Eq. (1) may be extended to

$$x_C^{(s)}(t) = \sum_{\alpha=1}^{m} C_\alpha f_\alpha^{(s)}(t) \qquad (3)$$

Obviously, the case $s = 0$ corresponds to the initial function, $s = -1$, to the integral etc. The error estimate of this value is

$$\sigma[x_C^{(s)}(t)] = \sum_{\alpha\beta=1}^{m} R_{\alpha\beta} f_\alpha^{(s)}(t) f_\beta^{(s)}(t), \qquad (4)$$

Where $R_{\alpha\beta}$ is the covariation matrix of the statistical errors of the coefficients. Usually it is replaced by a simplified expression

$$R_{\alpha\beta} = \sigma_{0m}^2 A_{\alpha\beta}^{-1}, \qquad (5)$$

$$A_{\alpha\beta} = \sum_{jk=1}^{m} h_{jk} f_\alpha(t_j) f_\beta(t_k) \qquad (6)$$

This approximation (5) is valid, only if the matrix $h_{jk} = \sigma_0^2 \mu_{jk}^{-1}$, where $\mu_{jk}$ is the covariance matrix of errors of the observations and $\sigma_0^2$ is an arbitrary positive constant. Andronov (1997) studied even more complicated case, when $h_{jk}$ are dependent on the time shift (like in the wavelet analysis). The classical case $w_k = \sigma_0^2 / \sigma_k^2$ was described e.g. by Anderson (??), Andronov (1994, 2003), Mikulášek (2007, 2015).

Marsakova and Andronov (1996) proposed a method of asymptotic parabola. The idea is that the part of the light curve near an extremum is approximated by two straight lines, which are connected by a piece of the parabola in a way that the function and its first derivative are continuous.

Free parameters for this approximation are the transition points between the parabola and straight lines corresponding to the ascending and descending branches. Extreme cases are the ordinary parabola and a broken line. The remaining three parameters - the slope of the lines and the vertical offset defined by the method of least squares.

The extremum of the smoothing function is determined in an usual way, by solving equation for the first derivative: $x'_C(t) = 0$ analytically or numerically by determining a maximum or minimum at a discrete grid of arguments, and then using iterations

$$t_e := t_e - \frac{x'_C(t)}{x''_C(t)} \qquad (7)$$

with a corresponding statistical error estimate:

$$\sigma[t_e] = \frac{\sigma[x'_C(t_e)]}{|x''_C(t_e)|} \qquad (8)$$

(cf. Andronov 1994, Mikulášek 2007). Obviously, this estimate is valid for a parabolic shape of the extremum and is not valid for flat extrema (e.g. total eclipses in binary stars). To determine of the statistically optimal number of parameters, we have computed approximations for few values of $m=2...m_{max}$, where $m_{max}= 10$ or smaller, and used $m$, which corresponds to the minimum of $\sigma[t_e]$. The computations stopped, when the degree of degeneracy

$$\gamma = \det(\mathbf{A}) / \prod_{\alpha=1}^{m} A_{\alpha\alpha}$$

of the matrix of normal equation becomes smaller than the limiting value $\gamma_{min} = 10^{-9}$.

**3. Applications**

For an illustration, we have used a part of the light curve of the RV – type pulsating star AC Her. We have used the international AAVSO database (http://aavso.org). After filtering of the photometrical data for outliers and other filters, the total number of the observations $N$=23540. The detailed analysis is in preparation. We have chosen a smaller part well covered by the observations.

In Fig. 1, the approximation using the "asymptotic parabola" is shown for few dozens cycles.

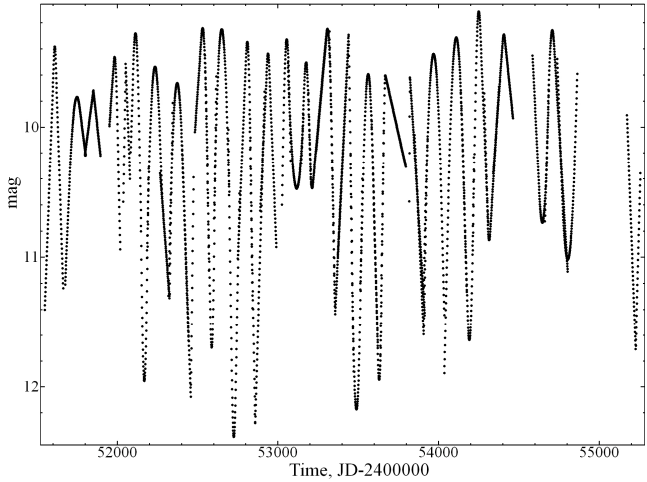

Figure 1: The "asymptotic parabola" approximation to the light curve of AC Her. The curve is compiled from the individual approximations of observations in the intervals surrounding maxima and minima.

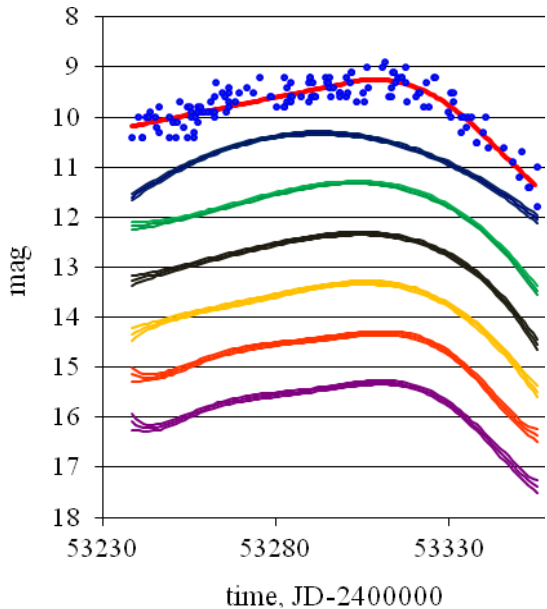

Figure 1: Original data (circles), approximations and "±1σ error corridor" (lines) for the asymptotic parabola (upper curve) and algebraic polynomials with number of parameters $m$=3..8.

An excellent approximation is seen at those parts where approximations near minimum and maximum overlap. Totally we determined characteristics of 749 extrema.

Fig 2 shows approximations using the "asymptotic parabola" and low-order algebraic polynomials of one of the extrema.

One may note that a maximum of the parabola is significantly shifted as compared to other approximations. However, the parabola corresponds to the best accuracy estimate of the time of extremum.

The dependences of the error estimates of the smoothing function are shown in Fig. 3. For this data set, the most accurate approximation corresponds to $m$=3 (parabola), the statistical errors increase with $m$.

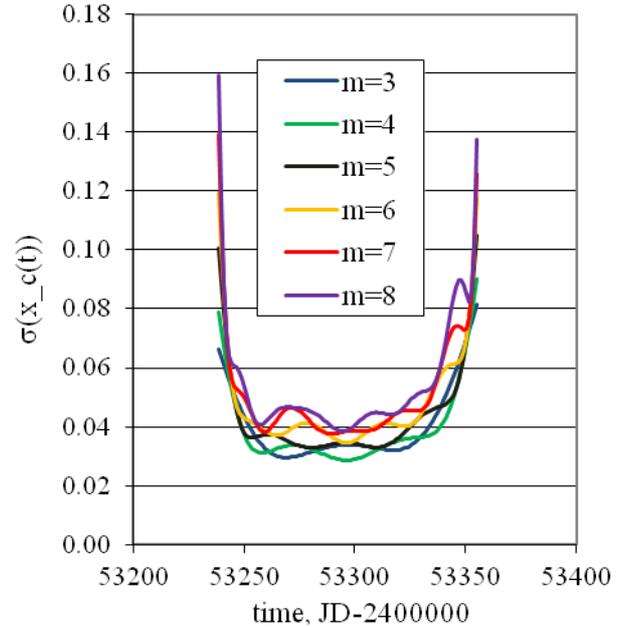

Figure 2: Scheme of internal and external contacts in eclipse.

## 4. Conclusions

The program was developed in the computer language VBA for Excel. It allows to determine the characteristics of extrema (maxima or minima) using the polynomial approximation with choosing a statistically optimal degree.

Results are applied to the pulsating variable S Aql.

The program will be used for further determination of extrema of variable stars of different types using our observations as well as observations from the virtual observatories and photometric databases like AAVSO, AFOEV, VSOLJ etc.